# FOUNDATIONS OF CLASSICAL ELECTRODYNAMICS, EQUIVALENCE PRINCIPLE AND COSMIC INTERACTIONS: A SHORT EXPOSITION AND AN UPDATE[*]

WEI-TOU NI,[†] HSIEN-HAO MEI,[‡] SHAN-JYUN WU[#]

*Center for Gravitation and Cosmology*
*Department of Physics, National Tsing Hua University*
*Hsinchu, Taiwan, 30013 Republic of China*

We look at the foundations of electromagnetism in this 1st LeCosPA Symposium. For doing this, after some review (constraints on photon mass etc.), we use two approaches. The first one is to formulate a Parametrized Post-Maxwellian (PPM) framework to include QED corrections and a pseudoscalar photon interaction. PPM framework includes lowest corrections to unified electromagnetism-gravity theories based on connection approach. It may also overlap with corrections implemented from generalized uncertainty principle (GUP) when electromagnetism-gravity coupling is considered. We discuss various vacuum birefringence experiments – ongoing and proposed -- to measure these parameters. The second approach -- the $\chi$-$g$ framework is to look at electromagnetism in gravity and various experiments and observations to determine its empirical foundation. The SME (Standard Model Extension) and SMS (Standard Model Supplement) overlap with the $\chi$-$g$ framework in their photon sector. We found that the foundation is solid with the only exception of a potentially possible pseudoscalar-photon interaction. We discussed its experimental constraints and look forward to more future experiments.

## 1 Introduction

### 1.1. *Classical Electrodynamics*

Classical electrodynamics is based on Maxwell equations and Lorentz force law. It can be derived by a least action with the following Lagrangian density for a system of charged particles in Gaussian units (e.g., Jackson [1]),

$$L_{EMS} = L_{EM} + L_{EM\text{-}P} + L_P = -(1/(16\pi))[(1/2)\eta^{ik}\eta^{jl} - (1/2)\eta^{il}\eta^{kj}]F_{ij}F_{kl} - A_k j^k - \Sigma_I m_I [(ds_I)/(dt)]\delta(\mathbf{x}-\mathbf{x}_I), \quad (1)$$

where $F_{ij} \equiv A_{j,i} - A_{i,j}$ is the electromagnetic field strength tensor with $A_i$ the electromagnetic 4-potential and comma denoting partial derivation, $\eta^{ij}$ is the Minkowskii metric with signature $(+, -, -, -)$, $m_I$ the mass of the $I$th charged particle, $s_I$ its 4-line element, and $j^k$ the charge 4-current density. Here, we use Einstein summation convention, i.e., summation over repeated indices. There are three terms in the Lagrangian density $L_{EMS}$ — (i) $L_{EM}$ for the electromagnetic field, (ii) $L_{EM\text{-}P}$ for the interaction of electromagnetic field and charged particles and (iii) $L_P$ for charged particles.

The electromagnetic field Lagrangian density (1) can be written in terms of the electric field **E** $[\equiv (E_1, E_2, E_3) \equiv (F_{01}, F_{02}, F_{03})]$ and magnetic induction **B** $[\equiv (B_1, B_2, B_3) \equiv (F_{32}, F_{13}, F_{21})]$ as

$$L_{EM} = (1/8\pi)[\mathbf{E}^2 - \mathbf{B}^2]. \tag{2}$$

**1.2. *Proca Lagrangian and the Photon Mass***

The classical Lagrangian density (1) is based on the photon having zero mass. To include the effects of nonvanishing photon mass $m_{photon}$, Proca (1936a, 1936b, 1936c, 1937, 1938) added a mass term $L_{Proca}$,

$$L_{Proca} = (m_{photon}^2 c^2 / 8\pi \hbar^2)(A_k A^k), \tag{3}$$

to the Lagrangian density of classical electrodynamics soon after Yukawa proposed short-range interaction in 1935. We use $\eta^{ij}$ and its inverse $\eta_{ij}$ to raise and lower indices. With this term, the Coulomb law is modified to have the electric potential $A_0$:

$$A_0 = q(e^{-\mu r}/r), \tag{4}$$

where $q$ is the charge of the source particle, $r$ is the distance to the source particle, and $\mu$ ($\equiv m_{photon} c/\hbar$) gives the inverse range of the interaction. The constraints on the mass and range of photons from various experiments are compiled in Table 1. For a comprehensive review, please see Goldhaber and Nieto (2010).

Table 1. Constraints on the mass and range of photon.

| Experiment/Observation | Mass constraint | Range constraint |
|---|---|---|
| Williams, Faller & Hill (1971): Lab Test | $m_{photon} \leq 10^{-14}$ eV ($= 2 \times 10^{-47}$ g) | $\mu^{-1} \geq 2 \times 10^7$ m |
| Davis, Goldhaber & Nieto (1975): Jupiter Magnetic field (Pioneer 10 Jupiter flyby) | $m_{photon} \leq 4 \times 10^{-16}$ eV ($= 7 \times 10^{-49}$ g) | $\mu^{-1} \geq 5 \times 10^8$ m |
| Ryutov (2007): Solar wind magnetic field | $m_{photon} \leq 10^{-18}$ eV ($= 2 \times 10^{-51}$ g) | $\mu^{-1} \geq 2 \times 10^{11}$ m |
| Chibisov (1976): Galactic sized mag. field | $m_{photon} \leq 2 \times 10^{-27}$ eV ($= 4 \times 10^{-60}$ g) | $\mu^{-1} \geq 10^{20}$ m |

As larger scale magnetic field discovered and measured, the constraints on photon mass and on the interaction range may become more stringent. If cosmic scale magnetic field is discovered, the constraint on the interaction range may become bigger or comparable to Hubble distance (of the order of radius of curvature of our observable universe). If this happens, the concept of photon mass may lose significance amid gravity coupling or curvature coupling of photons.

This paper is a short exposition of empirical foundations of electromagnetism with an update to include discussions of relevant recent theories and models. For a longer exposition, please see Ni (2012). The outline is as follows. In section 2, we present the Parametrized Post-Maxwell (PPM) framework for testing the foundations of classical electrodynamics in flat spacetime (including effective quantum corrections, but without gravity coupling), discuss its scope and summarize its usefulness. In section 3, we present the basic equations and discuss wave propagation in the PPM electrodynamics. In section 4, we discuss ultra-high precision laser interferometry experiments to measure the parameters of PPM electrodynamics. In section 5, we discuss empirical tests of electromagnetism in gravity and the $\chi$-g framework, and find pseudoscalar-photon interaction uniquely standing out. In section 6, we discuss the pseudoscalar-photon interaction, its relation to other approaches, and the use of radio galaxy observations and Cosmic Microwave Background (CMB) observations to constrain the cosmic polarization rotation induced by the pseudoscalar-photon interaction. In section 7, we present a summary and an outlook briefly.

## 2  Pamametrized Post-Maxwellian (PPM) Framework

For formulating a phenomenological framework for testing corrections to Maxwell-Lorentz classical electrodynamics, we notice that $(\mathbf{E}^2-\mathbf{B}^2)$ and $(\mathbf{E}\cdot\mathbf{B})$ are the only Lorentz invariants second order in the field strength, and $(\mathbf{E}^2-\mathbf{B}^2)^2$, $(\mathbf{E}\cdot\mathbf{B})^2$ and $(\mathbf{E}^2-\mathbf{B}^2)(\mathbf{E}\cdot\mathbf{B})$ are the only Lorentz invariants fourth order in the field strength. However, $(\mathbf{E}\cdot\mathbf{B})$ is a total divergence and, by itself in the Lagrangian density, does not contribute to the equation of motion (field equation). Multiplying $(\mathbf{E}\cdot\mathbf{B})$ by a pseudoscalar field $\varPhi$, the term $\varPhi(\mathbf{E}\cdot\mathbf{B})$ is the Lagrangian density for the pseudoscalar-photon (axion-photon) interaction. When this term is included together with the fourth-order invariants, we have the following phenomenological Lagrangian density for our Parametrized Post-Maxwell (PPM) Lagrangian density including various corrections and modifications to be tested by experiments and observations,

$$L_{PPM} = (1/8\pi)\{(\mathbf{E}^2-\mathbf{B}^2)+\xi\varPhi(\mathbf{E}\cdot\mathbf{B})+B_c^{-2}[\eta_1(\mathbf{E}^2-\mathbf{B}^2)^2+4\eta_2(\mathbf{E}\cdot\mathbf{B})^2+2\eta_3(\mathbf{E}^2-\mathbf{B}^2)(\mathbf{E}\cdot\mathbf{B})]\}, \qquad (5)$$

where

$$B_c \equiv E_c \equiv m^2c^3/e\hbar = 4.4\times10^{13}\text{ G} = 4.4\times10^{9}\text{ T} = 4.4\times10^{13}\text{ statvolt/cm} = 1.3\times10^{18}\text{ V/m}, \qquad (6)$$

with $e$ the absolute value of electron charge and $m$ the electron mass. This PPM Lagrangian density contains 4 parameters $\xi$, $\eta_1$, $\eta_2$ & $\eta_3$, and is an extension of the two-parameter ($\eta_1$ and $\eta_2$) post-Maxwellian Lagrangian density of Denisov, Krivchenkov and Kravtsov (2004). If there are absorptions, e.g., due to pair production or conversion to other particles, there would be imaginary part of the Lagrangian density. For example, one could add $L_{PPM}^{(Im)}$ to the Lagrangian density (5):

$$L_{PPM}^{(Im)} = (i/8\pi)\{ B_c^{-2}[\zeta_1(\mathbf{E}^2-\mathbf{B}^2)^2+4\zeta_2(\mathbf{E}\cdot\mathbf{B})^2+2\zeta_3(\mathbf{E}^2-\mathbf{B}^2)(\mathbf{E}\cdot\mathbf{B})]\}. \tag{7}$$

In this exposition, we are mainly concerned ourselves with the real part (5). To test the imaginary part (7), one may look into strong field pair production (e.g., Kim, 2011a, 2011b) and astrophysical phenomenon in strong field (e.g., Ruffini, Vereshchagin and Xue, 2010). In the Ruffini-Vereshchagin-Xue (2010) review of astrophysical phenomenon in strong field, their parameters, $\kappa_{2,0}$ and $\kappa_{2,1}$, corresponds to $\eta_1 = 8\pi B_c^2 \kappa_{2,0}$ and $\eta_2 = 2\pi B_c^2 \kappa_{0,2}$ in (5) and (7). In passing, we have noticed that in this first LeCosPA Symposium, there are talks related to pair productions and quantum fluctuations on acceleration and temperature (Labun and Rafelski, 2012: Unruh, 1976; S.Weinfurtner *et al.*, 2011) which could be subjected to similar kind of tests.

The manifestly Lorentz covariant form of Eq. (5) is

$$L_{PPM} = (1/(32\pi))\{-2F^{kl}F_{kl} -\xi\Phi F^{*kl}F_{kl}+B_c^{-2}[\eta_1(F^{kl}F_{kl})^2+\eta_2(F^{*kl}F_{kl})^2+\eta_3(F^{kl}F_{kl})(F^{*ij}F_{ij})]\}, \tag{8}$$

where

$$F^{*ij} \equiv (1/2)e^{ijkl} F_{kl}, \tag{9}$$

with $e^{ijkl}$ defined as

$$e^{ijkl} \equiv 1 \text{ if } (ijkl) \text{ is an even permutation of } (0123); -1 \text{ if odd; } 0 \text{ otherwise.} \tag{10}$$

Heisenberg-Euler (1936) Lagrangian density including the leading order quantum effects in slowly varying electric and magnetic field

$$L_{Heisenberg-Euler} = [2\alpha^2\hbar^2/(45(4\pi)^2m^4c^6)][(\mathbf{E}^2-\mathbf{B}^2)^2 + 7(\mathbf{E}\cdot\mathbf{B})^2], \tag{11}$$

fits the PPM framework with

$$\eta_1 = \alpha/(45\pi) = 5.1\times10^{-5}, \eta_2 = 7\alpha/(180\pi) = 9.0 \times10^{-5},\ \eta_3 = 0 \text{ and } \xi = 0, \tag{12}$$

where α is the fine structure constant.

Before Heisenberg & Euler (1936), Born and Infeld (Born, 1934; Born & Infeld, 1934) proposed the following (classical) Lagrangian density for the electromagnetic field

$$L_{Born-Infeld} = - (b^2/4\pi) [1 - (\mathbf{E}^2-\mathbf{B}^2)/b^2 - (\mathbf{E}\cdot\mathbf{B})^2/b^4]^{1/2}, \tag{13}$$

where $b$ is a constant which gives the maximum electric field strength. For field strength small compared with $b$, (13) can be expanded into

$$L_{Born-Infeld} = (1/8\pi) [(\mathbf{E}^2-\mathbf{B}^2) + (\mathbf{E}^2-\mathbf{B}^2)^2/b^2 + (\mathbf{E}\cdot\mathbf{B})^2/b^2 + O(b^{-4})]. \tag{14}$$

The lowest order of Born-Infeld electrodynamics agrees with the classical electrodynamics. The next order corrections fit the PPM framework Eq. (5) with

$$\eta_1 = \eta_2 = B_c^2/b^2, \text{ and } \eta_3 = \xi = 0. \tag{15}$$

In the Born-Infeld electrodynamics, $b$ is the maximum electric field. Electric fields at the edge of heavy nuclei are of the order of $10^{21}$ V/m. If we take $b$ to be $10^{21}$ V/m, then, $\eta_1 = \eta_2 = 5.9 \times 10^{-6}$.

*The PPM framework is useful in testing various models and theories of both electromagnetism and gravity.* A class of unified theories of electromagnetism and gravity with Lagrangian of the BF type (F: Curvature of the connection 1-form A (ω), with the gauge group U(2) (complexified) and with a potential for the B (Σ) field (Lie-algebra valued 2-form)) is proposed by Torres-Gomez, Krasnov and Scarinci (2010). Given a choice of a potential function with parameters α, γ, χ, δ and ξ, the theory is a deformation of (complex) general relativity and electromagnetism. With the reality conditions and using their equations (37), (38), (44), (45), the quadratic order plus quartic order Lagrangian can be put into the following form:

$$L^{(2)} + L^{(4)} = \alpha/(\gamma(\alpha+\gamma))\{ (\mathbf{E}^2 - \mathbf{B}^2) + (1/2)[\chi/\alpha(\alpha+\gamma)^3 + (2\delta/(\alpha\gamma(\alpha+\gamma)) + \xi(\alpha+\gamma)/\alpha\gamma^3)(\mathbf{E}^2 - \mathbf{B}^2)^2$$
$$- 2[\chi/\alpha(\alpha+\gamma)^3 - 2\delta/(\alpha\gamma(\alpha+\gamma)) + \xi(\alpha+\gamma)/\alpha\gamma^3] (\mathbf{E} \cdot \mathbf{B})^2$$
$$- 8i(\chi/\alpha(\alpha+\gamma)^3 - \xi(\alpha+\gamma)/\alpha\gamma^3) (\mathbf{E}^2 - \mathbf{B}^2)(\mathbf{E} \cdot \mathbf{B})]\}. \tag{16}$$

Comparing with (5) and (8), we have

$$\eta_1 = (1/2)B_c^2[\gamma\chi/\alpha(\alpha+\gamma)^3 + 2\delta/(\alpha\gamma(\alpha+\gamma)) + \xi(\alpha+\gamma)/\alpha\gamma^3], \eta_3 = \xi = 0,$$

$$\eta_2 = -(1/2)B_c^2[\gamma\chi/\alpha(\alpha+\gamma)^3 - 2\delta/(\alpha\gamma(\alpha+\gamma)) + \xi(\alpha+\gamma)/\alpha\gamma^3], \zeta_3 = -4B_c^2[\gamma\chi/\alpha(\alpha+\gamma)^3 - \xi(\alpha+\gamma)/\alpha\gamma^3]. \tag{17}$$

Thus, we see that experiments to measure the PPM parameters will also constrain the parameters of the proposed nonlinear electrodynamics from a class of unified theory of electromagnetism and gravity.

A focus in this Symposium is the Generalized Uncertainty Principle (GUP) as advocated by Bernard Carr (Carr *et al.*, 2011) and Pisin Chen (Chen and Wang, 2011). GUP affects the black hole entropy and the associated quantum effects in entropic gravity modify the Newton's gravitational law (Chen and Wang, 2011). Although the modification of gravity law is small, when the coupling to electromagnetism is considered/integrated/unified, the quartic corrections in the Lagrangian might not be negligible and, therefore, might be detectable by experiments to measure the PPM parameters.

In section 4, we will discuss how to measure the PPM parameters using birefringence measurements after we give the basic equations and discuss wave propagation in the PPM electrodynamics in section 3 in the following.

## 3 Basic Equations and Wave Propagation in the PPM Electrodynamics

In analogue with the nonlinear electrodynamics of continuous media, we can define the electric displacement **D** and magnetic field **H** as follows:

$$\mathbf{D} \equiv 4\pi(\partial L_{PPM}/\partial \mathbf{E}) = [1+2\eta_1(\mathbf{E}^2-\mathbf{B}^2)B_c^{-2}+2\eta_3(\mathbf{E}\cdot\mathbf{B})B_c^{-2}]\mathbf{E}+[\Phi+4\eta_2(\mathbf{E}\cdot\mathbf{B})B_c^{-2}+\eta_3(\mathbf{E}^2-\mathbf{B}^2)B_c^{-2}]\mathbf{B}, \quad (18)$$

$$\mathbf{H} \equiv -4\pi(\partial L_{PPM}/\partial \mathbf{B}) = [1+2\eta_1(\mathbf{E}^2-\mathbf{B}^2)B_c^{-2}+2\eta_3(\mathbf{E}\cdot\mathbf{B})B_c^{-2}]\mathbf{B}-[\Phi+4\eta_2(\mathbf{E}\cdot\mathbf{B})B_c^{-2}+\eta_3(\mathbf{E}^2-\mathbf{B}^2)B_c^{-2}]\mathbf{E}. \quad (19)$$

From **D** & **H**, we can define a second-rank $G_{ij}$ tensor, just like from **E** & **B** to define $F_{ij}$ tensor. With these definitions and following the standard procedure in electrodynamics [see, e.g., Jackson (1999), p. 599], the nonlinear equations of the electromagnetic field are

$$\text{curl } \mathbf{H} = (1/c)\, \partial \mathbf{D}/\partial t + 4\pi\, \mathbf{J}, \quad (20)$$

$$\text{div } \mathbf{D} = 4\pi\, \rho, \quad (21)$$

$$\text{curl } \mathbf{E} = -(1/c)\, \partial \mathbf{B}/\partial t, \quad (22)$$

$$\text{div } \mathbf{B} = 0. \quad (23)$$

We notice that it has the same form as in macroscopic electrodynamics. The Lorentz force law remains the same as in classical electrodynamics:

$$d[(1-\mathbf{v}_I^2/c^2)^{-1/2}m_I\mathbf{v}_I]/dt = q_I[\mathbf{E} + (1/c)\mathbf{v}_I \times \mathbf{B}], \quad (24)$$

for the $I$-th particle with charge $q_I$ and velocity $\mathbf{v}_I$ in the system. The source of $\Phi$ in this system is $(\mathbf{E}\cdot\mathbf{B})$ and the field equation for $\Phi$ is

$$\partial^i L_\Phi/\partial(\partial^i \Phi) - \partial L_\Phi/\partial \Phi = \mathbf{E}\cdot\mathbf{B}, \quad (25)$$

where $L_\Phi$ is the Lagrangian density of the pseudoscalar field $\Phi$.

Following the previous method (Ni *et al.*, 1991; Ni, 1998; Ni, 2012), i.e., separating the electric field **E** and magnetic induction field **B** into the wave part $\mathbf{E}^{wave}$, $\mathbf{B}^{wave}$ (small compared to external part) and external part $\mathbf{E}^{ext}$, $\mathbf{B}^{ext}$, and linearizing the equations of motion, one can derive the PPM wave propagation equations and obtain the dispersion relations (Ni, 2012). From the dispersion relations, the principal indices of refraction can be found. The necessary and sufficient conditions of "no birefringence" on the PPM parameters are

$$\eta_1 = \eta_2, \eta_3 = 0, \text{ and no constraint on } \xi. \quad (26)$$

The Born-Infeld electrodynamics satisfies this condition and has no birefringence in the theory.

For $\mathbf{E}^{ext} = 0$, the (principal) refractive indices in the transverse external magnetic field $\mathbf{B}^{ext}$ for the linearly polarized lights whose polarizations are parallel and orthogonal to the magnetic field, are as follows:

$$n_\parallel = 1 + \{(\eta_1+\eta_2) + [(\eta_1-\eta_2)^2 +\eta_3^2]^{1/2}\} (\mathbf{B}^{ext})^2 B_c^{-2} \quad (\mathbf{E}^{wave} \parallel \mathbf{B}^{ext}), \tag{27}$$

$$n_\perp = 1 + \{(\eta_1+\eta_2) - [(\eta_1-\eta_2)^2 +\eta_3^2]^{1/2}\} (\mathbf{B}^{ext})^2 B_c^{-2} \quad (\mathbf{E}^{wave} \perp \mathbf{B}^{ext}). \tag{28}$$

For $\mathbf{B}^{ext} = 0$, the (principal) refractive indices in the transverse external electric field $\mathbf{E}^{ext}$ for the linearly polarized lights whose polarizations are parallel and orthogonal to the magnetic field, are as follows:

$$n_\parallel = 1 + \{(\eta_1+\eta_2) + [(\eta_1-\eta_2)^2 +\eta_3^2]^{1/2}\} (\mathbf{E}^{ext})^2 B_c^{-2} \quad (\mathbf{E}^{wave} \parallel \mathbf{E}^{ext}), \tag{29}$$

$$n_\perp = 1 + \{(\eta_1+\eta_2) - [(\eta_1-\eta_2)^2 +\eta_3^2]^{1/2}\} (\mathbf{E}^{ext})^2 B_c^{-2} \quad (\mathbf{E}^{wave} \perp \mathbf{E}^{ext}). \tag{30}$$

The magnetic field near pulsars can reach $10^{12}$ G, while the magnetic field near magnetars can reach $10^{15}$ G. The astrophysical processes in these locations need nonlinear electrodynamics to model. In the following section, we turn to experiments to measure the parameters of the PPM electrodynamics.

## 4. Measuring the parameters of the PPM electrodynamics

There are four parameters $\eta_1$, $\eta_2$, $\eta_3$, and $\xi$ in PPM electrodynamics to be measured by experiments. For the QED (Quantum Electrodynamics) corrections to classical electrodynamics, $\eta_1 = \alpha/(45\pi) = 5.1 \times 10^{-5}$, $\eta_2 = 7\alpha/(180\pi) = 9.0 \times 10^{-5}$, $\eta_3 = 0$, and $\xi = 0$. There are three vacuum birefringence experiments on going in the world to measure this QED vacuum birefringence – the BMV experiment (Battesti *et al.*, 2008), the PVLAS experiment (Zavattini *et al.*. 2008) and the Q & A (QED vacuum birefringence and Axion search) experiment (Chen *et al.*, 2007; Mei *et al.*, 2010). The QED vacuum birefringence $\Delta n$ in a magnetic field $\mathbf{B}^{ext}$ is

$$\Delta n = n_\parallel - n_\perp = 4.0 \times 10^{-24} (\mathbf{B}^{ext}/1T)^2. \tag{31}$$

For 2.3 T field of the Q & A rotating permanent magnet, $\Delta n$ is $2.1 \times 10^{-23}$. This is about the same order of magnitude change in fractional length that ground interferometers for gravitational-wave detection aim at. Quite a lot of techniques developed in the gravitational-wave detection community are readily applicable to vacuum birefringence measurement (Ni *et al.*, 1991).

The basic principle of these experimental measurements is shown as Figure 1. The laser light goes through a polarizer and becomes polarized. This polarized light goes through a region of magnetic field. Its polarization status is subsequently analyzed by the analyzer-detector subsystem to extract the polarization effect imprinted in the region of the magnetic field. In the actual experiments, one has to multiply the optical pass through the magnetic field by using reflections or Fabry-Perot cavities.

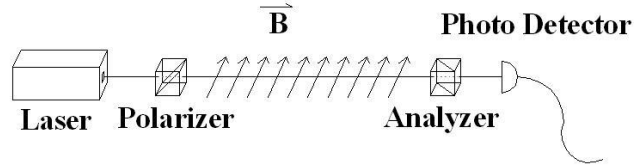

Figure 1. Principle of vacuum birefringence and dichroism measurement.

For our Q & A experiment, the facility is shown in Figure 2. Photo on the left shows the Q & A apparatus for Phase II experiment (Chen *et al*., 2007); photo in the middle shows the Q & A apparatus for Phase III experiment (Mei *et al*., 2010); the upper right photo shows the mirror suspension; the lower right photo shows the laser injection table. Two vacuum tanks shown on the left photo of Figure 2 house two 5 cm-diameter Fabry-Perot mirrors with their suspensions respectively; the 0.6 m 2.3 T permanent magnet is between two tanks. For Phase III, we double the distance of two Fabry-Perot mirrors to 7 m, and insert another 2.3 T permanent magnet with magnetic field length 1.8 m rotatable up to 13 cycle/s.

All three ongoing experiments – PVLAS, Q & A, and BMV – are measuring the birefringence $\Delta n$, and hence, $\eta_1 - \eta_2$ in case $\eta_3$ is assumed to be zero. To measure $\eta_1$ and $\eta_2$ separately, one-arm common path polarization measurement interferometer is not enough. We need a two-arm interferometer with the paths in two arms in magnetic fields with different strengths (or one with no magnetic field).

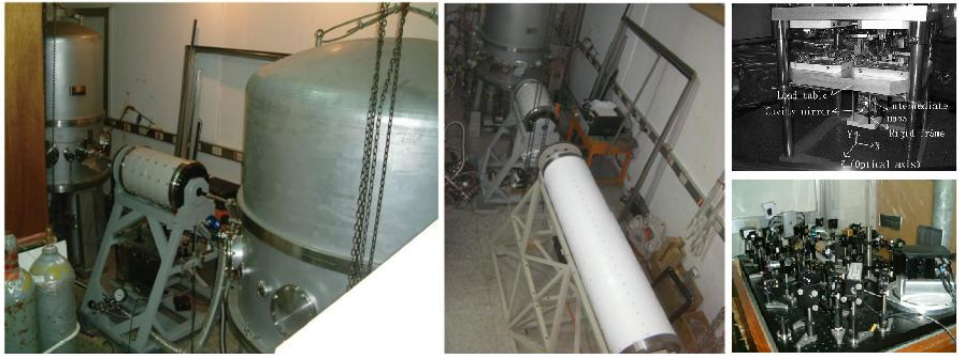

Figure 2. Photo on the left shows the Q & A apparatus for Phase II experiment; photo in the middle shows the Q & A apparatus for Phase III experiment; the upper right photo shows a mirror suspension; the lower right photo shows the laser injection table.

To measure $\eta_3$ in addition, one needs to use both external electric and external magnetic field. One possibility is to let light goes through strong microwave cavity and interferes (Ni, 2012).

As to the term $\xi\Phi$ and parameter $\xi$, it does not give any change in the index of refraction. However, it gives a polarization rotation and the effect can be measured though observations with astrophysical and cosmological propagation of electromagnetic waves (Section 6).

## 5. Empirical tests of electromagnetism in gravity and the $\chi$-$g$ framework

In section 1, we have discussed the constraints on Proca part of Lagrangian density, i.e., photon mass. In this section, we discuss the empirical foundation of the Maxwell (main) part of electromagnetism. Since gravity is everywhere, for doing this, we use the $\chi$-$g$ framework (Ni, 1983, 1984a, 1984b, 2010) which is summarized in the following interaction Lagrangian density

$$L_I = -(1/(16\pi))\chi^{ijkl} F_{ij} F_{kl} - A_k j^k (-g)^{(1/2)} - \Sigma_I m_I (ds_I)/(dt)\, \delta(\mathbf{x}-\mathbf{x}_I), \tag{32}$$

with $\chi^{ijkl} = \chi^{klij} = -\chi^{jikl}$ a tensor density of the gravitational fields (e.g., $g_{ij}$, $\varphi$, etc.) or fields to be investigated. The gravitational constitutive tensor density $\chi^{ijkl}$ dictates the behaviour of electromagnetism in a gravitational field and has 21 independent components in general. For general relativity or a metric theory (when EEP holds), $\chi^{ijkl}$ is determined completely by the metric $g_{ij}$ and equals $(-g)^{1/2}[(1/2)g^{ik}g^{jl}-(1/2)g^{il}g^{jk}]$; when $g^{ik}$ is replaced by $\eta^{ik}$, we obtain the special relativistic Lagrangian density (1). The SME (Standard Model Extension; Kostelecky and Mews, 2002) and SMS (Standard Model Supplement; Zhou and Ma, 2010, 2011; Ma 2012) overlap the $\chi$-$g$ framework in their photon sector. Hence, our studies are directly relevant to parameter constraints in these models.

In the following, we summarize experimental constraints on the 21 degrees of freedom of $\chi^{ijkl}$ to see how close we can reach EEP and metric theory empirically. This procedure also serves to reinforce the empirical foundations of classical electromagnetism as EEP locally is based on special relativity including classical electromagnetism. For a more detailed survey, see Ni (2012) and references therein.

*Constraints from no birefringence*: In the $\chi$-$g$ framework, the theoretical condition for no birefringence (no splitting, no retardation) for electromagnetic wave propagation in all directions is that the constitutive tensor $\chi^{ijkl}$ can be written in the following form

$$\chi^{ijkl}=(-H)^{1/2}[(1/2)H^{ik} H^{jl}-(1/2)H^{il} H^{kj}]\psi + \varphi e^{ijkl}, \tag{33}$$

where $H = \det(H_{ij})$ and $H_{ij}$ is a metric which generates the light cone for electromagnetic propagation (Ni, 1983, 1984a,b; Lämmerzahl and Hehl 2004). Polarization measurements of light from pulsars and cosmologically distant astrophysical sources yield stringent constraints agreeing with (33) down to $2 \times 10^{-32}$ fractionally; for a review, see Ni (2010).

In the remaining part of this section, we assume (33) to be valid. Note that (33) has an axion degree of freedom, $\varphi e^{ijkl}$, and a 'dilaton' degree of freedom, $\psi$. To fully recover EEP, we still need (i) good constraints on only one physical metric, (ii) good constraints on no $\psi$ ('dilaton'), and (iii) good constraints on no $\varphi$ (axion) or no pseudoscalar-photon interaction.

*Constraints on one physical metric and no 'dilaton' ($\psi$)*: Let us now look into the empirical constraints for $H^{ij}$ and $\psi$. In Eq. (32), $ds$ is the line element determined from the metric $g_{ij}$. From Eq. (33), the gravitational coupling to electromagnetism is determined by the metric $H_{ij}$ and two (pseudo)scalar fields $\varphi$ 'axion' and $\psi$ 'dilaton'. If $H_{ij}$ is not proportional to $g_{ij}$, then the hyperfine levels of the lithium atom, the beryllium atom, the mercury atom and other atoms will have additional shifts. But this is not observed to high accuracy in Hughes-Drever-type experiments. Therefore $H_{ij}$ is proportional to $g_{ij}$ to certain accuracy. Since a change of $H^{ik}$ to $\lambda H^{ij}$ does not affect $\chi^{ijkl}$ in Eq. (33), we can define $H_{11} = g_{11}$ to remove this scale freedom (Ni, 1983, 1984a). For a review, see Ni (2010).

Eötvös-Dicke experiments (Schlamminger *et al.*, 2008 and references therein) are performed on unpolarized test bodies. In essence, these experiments show that unpolarized electric and magnetic energies follow the same trajectories as other forms of energy to certain accuracy. The constraints on Eq. (33) are

$$| 1-\psi | / U < 10^{-10}, \tag{34}$$

and

$$| H_{00} - g_{00} | / U < 10^{-6}, \tag{35}$$

where $U$ ($\sim 10^{-6}$) is the solar gravitational potential at the earth.

In 1976, Vessot *et al.* (1980) used an atomic hydrogen maser clock in a space probe to test and confirm the metric gravitational redshift to an accuracy of $1.4 \times 10^{-4}$, i.e.,

$$| H_{00} - g_{00} | / U \leq 1.4 \times 10^{-4}, \tag{36}$$

where $U$ is the change of earth gravitational field that the maser clock experienced.

With constraints from (i) no birefringence, (ii) no extra physical metric, (iii) no $\psi$ ('dilaton'), we arrive at the theory (32) with $\chi^{ijkl}$ given by

$$\chi^{ijkl} = (-g)^{1/2} [(1/2)\, g^{ik} g^{jl} - (1/2)\, g^{il} g^{kj} + \varphi\, \varepsilon^{ijkl}], \tag{37}$$

i.e., an axion theory (Ni, 1983, 1984a; Hehl and Obukhov 2008). Here $\varepsilon^{ijkl}$ is defined to be $(-g)^{-1/2}\, e^{ijkl}$. The current constraints on $\varphi$ from astrophysical observations and CMB polarization observations will be discussed in the next section. Thus, from experiments and observations, only one degree of freedom of $\chi^{ijkl}$ is not much constrained.

Now let's turn into more formal aspects of equivalence principles. We proved that for a system whose Lagrangian density given by equation (32), the Galileo Equivalence Principle (UFF [Universality of Free Fall; WEP I) holds if and only if equation (37) holds (Ni, 1974, 1977).

If $\varphi \neq 0$ in (37), the gravitational coupling to electromagnetism is not minimal and EEP is violated. Hence WEP I does not imply EEP and Schiff's conjecture (which states that WEP I implies EEP) is incorrect (Ni, 1973, 1974, 1977). However, WEP I does constrain the 21 degrees of freedom of $\chi$ to only one degree of freedom ($\varphi$), and Schiff's conjecture is largely right in spirit.

The theory with $\varphi \neq 0$ is a pseudoscalar theory with important astrophysical and cosmological consequences (section 6). This is an example that investigations in fundamental physical laws lead to implications in cosmology (Ni, 1977). Investigations of CP problems in high energy physics leads to a theory with a similar piece of Lagrangian with $\varphi$ the axion field for QCD (Peccei and Quinn, 1977; Weinberg, 1978; Wilczek, 1978).

In this section, we have shown that the empirical foundation of classical electromagnetism is solid except in the aspect of a possible pseudoscalar photon interaction. This exception has important consequences in cosmology. In the following section, we address this issue.

**6. Pseudoscalar-photon interaction**

In this section, we discuss the modified electromagnetism in gravity with the pseudoscalar-photon interaction which we have reached in the last section, i.e., the theory (32) with the constitutive tensor density (33). Its Lagrangian density is as follows

$$L_I = -(1/(16\pi))(-g)^{1/2}[(1/2)g^{ik}g^{jl}-(1/2)g^{il}g^{kj}+\varphi\, \varepsilon^{ijkl}]F_{ij}F_{kl} - A_k j^k (-g)^{(1/2)} - \Sigma_I m_I (ds_I)/(dt)\delta(\mathbf{x}-\mathbf{x}_I). \tag{38}$$

In the constitutive tensor density and the Lagrangian density, $\varphi$ is a scalar or pseudoscalar function of relevant variables. If we assume that the $\varphi$-term is local CPT invariant, than $\varphi$ should be a pseudoscalar (function) since $\varepsilon^{ijkl}$ is a pseudotensor. The pseudoscalar(scalar)-photon interaction part (or the nonmetric part) of the Lagrangian density of this theory is

$$L^{(\varphi\gamma\gamma)} = L^{(NM)} = -(1/16\pi)\, \varphi\, e^{ijkl}F_{ij}F_{kl} = -(1/4\pi)\, \varphi_{,i}\, e^{ijkl}A_j A_{k,l} \text{ (mod div)}, \tag{39}$$

where 'mod div' means that the two Lagrangian densities are related by integration by parts in the action integral. This term gives pseudoscalar-photon-photon interaction in the quantum regime and can be denoted by $L^{(\varphi\gamma\gamma)}$. *This term is also the $\xi$-term in the PPM Lagrangian density $L_{PPM}$ with the $\varphi \equiv (1/4)\xi\Phi$ correspondence.* The modified Maxwell equations (Ni 1973, 1977) from Eq. (38) are

$$F^{ik}{}_{;k} + \varepsilon^{ikml} F_{km}\varphi_{,l} = -4\pi j^i, \tag{40}$$

where the covariant derivation ; is with respect to the Christoffel connection of the metric. The Lorentz force law is the same as in metric theories of gravity or general relativity. Gauge invariance and charge conservation are guaranteed. The modified Maxwell equations are also conformally invariant.

The rightest term in equation (39) is reminiscent of Chern-Simons (1974) term $e^{\alpha\beta\gamma} A_\alpha F_{\beta\gamma}$. There are two differences: (i) Chern-Simons term is in 3 dimensional space; (ii) Chern-Simons term as integrand in the integral is a total divergence (Table 2).

**Table 2.** Various terms in the Lagrangian and their meanings.

| Term | Dimension | Reference | Meaning |
|---|---|---|---|
| $e^{\alpha\beta\gamma} A_\alpha F_{\beta\gamma}$ | 3 | Chern-Simons (1974) | Integrand for topological invariant |
| $e^{ijkl} \varphi F_{ij} F_{kl}$ | 4 | Ni (1973, 1974, 1977) | Pseudoscalar-photon coupling |
| $e^{ijkl} \varphi F^{QCD}{}_{ij} F^{QCD}{}_{kl}$ | 4 | Peccei-Quinn (1977) Weinberg (1978) Wilczek (1978) | Pseudoscalar-gluon coupling |
| $e^{ijkl} V_i A_j F_{kl}$ | 4 | Carroll-Field-Jackiw (1990) | External constant vector coupling |

A term similar to the one in equation (39) (axion-gluon interaction) occurs in QCD in an effort to solve the strong CP problem (Peccei and Quinn, 1977; Weinberg, 1978; Wilczek, 1978). Carroll, Field and Jackiw (1990) proposed a modification of electrodynamics with an additional $e^{ijkl} V_i A_j F_{kl}$ term with $V_i$ a constant vector (See also Jackiw, 2007). This term is a special case of the term $e^{ijkl} \varphi F_{ij} F_{kl}$ (mod div) with $\varphi_{,i} = -½ V_i$. Various terms discussed are listed in Table 2.

Polarization rotation is induced in the propagation of linearly polarized electromagnetic wave obeying the modified Maxwell equations (40) in $\varphi$-field. This rotation in the long range propagation in cosmos is called cosmic polarization rotation. Empirical tests/constraints of the pseudoscalar-photon interaction come from polarization observations of radio and optical/UV polarization of radio galaxies, and of cosmic microwave background (CMB). The constraints obtained from these observations on the cosmic polarization rotation angle $\Delta\varphi$ are within ± 30 mrad. Converting to constraints on $\xi$ and $\Delta\Psi$, we have $|\xi\Delta\Psi| = ± 0.12$. (Ni, 2012; and references therein).

## 7. Outlook

We have looked at the foundations of electromagnetism in this short exposition. For doing this, we have used two approaches. The first one is to formulate a Parametrized Post-Maxwellian framework to include QED corrections and a pseudoscalar photon interaction. We discuss various vacuum birefringence experiments — ongoing and

proposed — to measure these parameters. The second approach is to look at electromagnetism in gravity and various experiments and observations to determine its empirical foundation. We found that the foundation of EEP of the gravity coupling to classical electrodynamics is solid with the only exception of a potentially possible pseudoscalar-photon interaction. This provides the empirical foundation for our first approach to include quantum corrections, possible unification modifications and pseudoscalar-photon interaction. We have discussed various experimental constraints and look forward to more future experiments.

**Acknowledgments**

We would like to thank Bernard Carr, Pisin Chen, Dah-Wei Chiou, Sang Pyo Kim, Lance Laben, Bo-Qiang Ma, Chiao-Hsuan Wang and She-Sheng Xue for helpful discussions. We would also like to thank the National Science Council (Grants No. NSC100-2119-M-007-008 and No. NSC100-2738-M-007-004) for supporting this work in part. One of us (WTN) would like to thank Leung Center for Cosmology and Particle Astrophysics (LeCosPA Center) for invitation to the First LeCosPA Symposium.

**References**

R. Battesti *et al*. (BMV Collaboration) (2008), The BMV Experiment: a Novel Apparatus to Study the Propagation of Light in a Transverse Magnetic Field, *Eur. Phys. J. D* **46**, 323-333.

M. Born (1934), *Proc. R. Soc. London,* **A143**, 410.

M. Born and L. Infeld (1934), *Proc. R. Soc. London,* **A144**, 425.

B. Carr, L. Modesto and I. Prémont-Schwarz (2011), Generalized Uncertainty Principle and Self-dual Black Holes, arXiv:1107.0708.

S. M. Carroll, G. B. Field and R. Jackiw (1990), Limits on a Lorentz- and parity-violating modification of electrodynamics, *Phys. Rev.* **D41**, 1231-1240.

P. Chen, C.-H. Wang (2011), Where is hbar Hiding in Entropic Gravity? arXiv:1112.3078.

S.-J. Chen, H.-H. Mei and Ni, W.-T. (Q & A Collaboration) (2007), Q & A Experiment to Search for Vacuum Dichroism, Pseudoscalar Photon Interaction and Millicharged Fermions. *Mod. Phys. Lett. A* **22**, 2815-2831 [arXiv:hep-ex/0611050].

S.-S. Chern, and J. Simons (1974), Characteristic forms and geometric invariants, *The Annals of Mathematics*, 2$^{nd}$ Ser. **99**, 48.

G. V. Chibisov (1976), Astrophysical upper limits on the photon rest mass, *Uspekhi Fizicheskikh Nauk*, **119**, 551–555 [*Soviet Physics Uspekhi*, **19**, 624–626 (1976)].

L. Davis, Jr., A. S. Goldhaber and M. M. Nieto (1975), Limit on the photon mass deduced from Pioneer-10 observations of Jupiter's magnetic field, *Phys. Rev. Lett.,* **35**, 1402–1405.

V. I. Denisov, I. V.Krivchenkov and N. V. Kravtsov (2004), Experiment for Measuring the post-Maxwellian Parameters of Nonlinear Electrodynamics of Vacuum with Laser-Interferometer Techniques, *Phys. Rev.* **D69**, 066008.

A. S. Goldhaber and M. M. Nieto (2010), Photon and Graviton Mass Limits, *Rev. Mod. Phys.* **82**, 939-979.

F. W. Hehl and Yu. N. Obukhov (2008), Equivalence principle and electromagnetic field: no birefringence, no dilaton, and no axion, *Gen. Rel. Grav.,* **40**, 1239-1248.

W. Heisenberg and E. Euler (1936), *Zeitschrift für Physik*, **98**, 714.

R. Jackiw (2007), Lorentz Violation in a Diffeomorphism-Invariant Theory, *CPT'07 Proceedings* [arXiv: 0709.2348].

J. D. Jackson (1999), *Classical Electrodynamics*, John Wiley & Sons, Hoboken.

S. P. Kim (2011a), QED Effective Actions in Space-Dependent Gauge and Electromagnetic Duality, arXiv:1105.4382v2 [hep-th].

S. P. Kim (2011b), QED Effective Action in Magnetic field Backgrounds and Electromagnetic Duality, *Phys. Rev.* **D84**, 065004.

V. A. Kostelecky and M. Mewes 2002, Signals for Lorentz violation in Electrodynamics, *Phys. Rev.* **D66,** 056005.

L. Labun and J. Rafelski (2012), Acceleration and Vacuum Temperature, arXiv:1203.6148.

C. Lämmerzahl and F. W. Hehl (2004), Riemannian Light Cone from Vanishing Birefringence in Premetric Vacuum Electrodynamics, *Phys. Rev. D* **70**, 105022.

B.-Q. Ma, New Perspective on Space and Time from Lorentz Violation, arXiv:1203.5852 (2012).

H.-H. Mei, W.-T. Ni, S.-J. Chen and S.-s. Pan (Q & A Collaboration) (2010), Axion Search with Q & A Experiment, *Mod. Phys. Lett*. A 25, 983–993 [arXiv:1001.4325].

W.-T. Ni 1973 A Nonmetric Theory of Gravity, preprint, Montana State University [http://astrod.wikispaces.com/].

W.-T. Ni (1974), Weak equivalence principles and gravitational coupling, *Bull. Am. Phys. Soc.* **19**, 655.

W.-T. Ni (1977), Equivalence principles and electromagnetism, *Phys. Rev. Lett.* **38** 301–4.

W.-T. Ni (1983), Equivalence Principles, Their Empirical Foundations, and the Role of Precision Experiments to Test Them, *Proceedings of the 1983 International School and Symposium on Precision Measurement and Gravity Experiment, Taipei, Republic of China, January 24-February 2, 1983*, W.-T. Ni, (Ed.), (Published by National Tsing Hua University, Hsinchu, Taiwan, Republic of China), pp. 491-517 [http://astrod.wikispaces.com/].

W.-T. Ni (1984a), Equivalence Principles and Precision Experiments, *Precision Measurement and Fundamental Constants II*, B. N. Taylor and W. D. Phillips, (Ed.), Natl. Bur. Stand. (U S) Spec. Publ. **617,** pp 647-651.

W.-T. Ni, (1984b). Timing Observations of the Pulsar Propagations in the Galactic Gravitational Field as Precision Tests of the Einstein Equivalence Principle, *Proceedings of the Second Asian-Pacific Regional Meeting of the International Astronomical Union*, B. Hidayat and M. W. Feast (Ed.), (Published by Tira Pustaka, Jakarta, Indonesia), pp. 441-448

W.-T. Ni, K. Tsubono, N. Mio, K. Narihara, S.-C. Chen, S.-K. King and S.-s. Pan (1991), Test of Quantum Electrodynamics using Ultra-High Sensitive Interferometers, *Mod. Phys. Lett. A* **6**, 3671-3678.

W.-T. Ni (1998), Magnetic Birefringence of Vacuum—Q & A Experiment, *Frontier Test of QED and Physics of the Vacuum*, Eds. E. Zavattini, D. Bakalov, C. Rizzo, 1998, Heron Press, Sofia, pp. 83-97.

W.-T. Ni (2010), Searches for the Role of Spin and Polarization in Gravity, *Reports on Progress in Physics* **73**, 056901

W.-T. Ni (2012), Foundations of Electromagnetism, Equivalence Principles and Cosmic Interactions, Chaper 3 in *Trends in Electromagnetism - From Fundamentals to Applications*, pp. 45-68 (March, 2012), Victor Barsan (Ed.), ISBN: 978-953-51-0267-0, InTech (open access) [arXiv:1109.5501], Available from: http://www.intechopen.com/books/trends-in-electromagnetism-from-fundamentals-to-applications/foundations-of-electromagnetism-equivalence-principles-and-cosmic-interactions.

R. D. Peccei and H. R. Quinn (1977), CP Conservation in the presence of pseudoparticles, *Phys. Rev. Lett.* **38**,. 1440-1443.

A. Proca (1936a), Sur la théorie du positron, *Comptes Rendus de l'Académie des Sciences* **202**, 1366–1368.

A. Proca (1936b), Sur la théorie ondulatoire des electrons positifs et negatives, *Journal de Physique et Le Radium* **7**, 347–353.

A. Proca (1936c), Sur les photons et les particules charge pure, *Comptes Rendus de l'Académie des Sciences* **203**, 709–711.

A. Proca (1937), Particles libres: Photons et particules 'charge pure', *Journal de Physique et Le Radium* **8**, 23–28.

A. Proca (1938), Théorie non relativiste des particles a spin entire, *Journal de Physique et Le Radium*. **9**, 61–66.

R. Ruffini, G. Vereshchagin and S.-S. Xue (2010), Electron-Positron Pairs in Physics and Astrophysics: from Heavy Nuclei to Black Holes, *Phys. Reports*, **487**, 1-140.

D. D. Ryutov (2007), Using Plasma Physics to Weigh the Photon, *Plasma Physics and Controlled Fusion*, **49**, B429–B438.

S. Schlamminger, K.-Y. Choi, T. A. Wagner, J. H. Gundlach and E. G. Adelberger (2008), Test of the equivalence principle using a rotating torsion balance, *Phys. Rev. Lett.* **100**, 041101.

A. Torres-Gomez, K. Krasnov and C. Scarinci (2011), Unified Theory of Nonlinear Electrodynamics and Gravity, *Phys. Rev.* **D83**, 025023.

W. G. Unruh (1976), *Phys. Rev.* **D14**, 870.

R. F. C. Vessot *et al*. (1980), Test of relativistic gravitation with a space-borne hydrogen maser, *Phys. Rev. Lett*. **45**, 2081-2084.

S. Weinberg (1978), A New Light Boson? *Phys. Rev. Lett.* **40**, 233.

S. Weinfurtner *et al.* (2011), Measurement of Stimulated Hawking Emission in an Analogue System, *Phys. Rev. Lett.* **106**, 021302.

F. Wilczek (1978), Problem of strong P and T invariance in the presence of instantons, *Phys. Rev. Lett.* **40**, 279.

E. R. Williams, J. E. Faller and H. A. Hill (1971), New Experimental Test of Coulomb's Law: A Laboratory Upper Limit on the Photon Rest Mass, *Phys. Rev. Lett.*, **26**, 721–724.

E. Zavattini *et al*. (PVLAS Collaboration) (2008), New PVLAS Results and Limits on Magnetically Induced Optical Rotation and Ellipticity in Vacuum, *Phys. Rev. D* **77**, 032006.

L. Zhou and B.-Q. Ma, *Mod. Phys. Lett. A* **25,** 2489 (2010) [arXiv:1009.1331].

L. Zhou and B.-Q. Ma, *Chin. Phys. Lett. C* **35,** 987 (2011) [arXiv:1109.6387].